\documentstyle[prb,aps,amsfonts,amstex,epsfig,eqsecnum]{revtex}
\begin{document}
\title{Andreev scattering and Josephson current in a one-dimensional electron liquid}   

\author {Ian Affleck$^{1,2,3}$, Jean-S\'ebastien Caux$^2$ and Alexandre 
M. Zagoskin$^2$}

\address{$^1$Canadian Institute for Advanced Research, The Universtity of 
British Columbia, Vancouver, BC, V6T 1Z1, Canada \\
$^2$Department of Physics and Astronomy, The University of British Columbia, 
Vancouver, BC, V6T 1Z1, Canada\\
$^3$ Institute for Theoretical Physics, University of California,
Santa Barbara, CA93117-4030}
\date{$Date:$ \today}
\maketitle
\draft
\begin{abstract}
Andreev scattering and the Josephson current through a one-dimensional
interacting electron liquid  
sandwiched between two superconductors are re-examined.  We first
present some apparently new 
results on the non-interacting case by studying an exactly solvable tight-binding model 
rather than the usual
continuum model.  We show that perfect Andreev scattering (i.e. zero normal scattering) 
at the Fermi energy can only be achieved
by fine-tuning junction parameters, a fine-tuning which is possible
 even with bandwidth mismatch between superconductor and normal metal.
We also obtain exact results for the Josephson current, which is generally a smooth
function of the superconducting phase difference except when the
junction parameters are adjusted to give  
perfect Andreev scattering, in which case it becomes a sawtooth
function. We then observe that, even 
when interactions are included, all
low energy properties of a junction ($E<<\Delta$, the superconducting gap)
 can be obtained by ``integrating out'' the superconducting electrons
to obtain an effective Hamiltonian 
describing the metallic electrons only with a boundary pairing interaction.  This boundary model
provides a suitable starting point for bosonization/renormalization group/boundary conformal field
theory analysis.  We argue that  total normal reflection and total Andreev reflection correspond to
two fixed points of the boundary renormalization group.  For repulsive bulk interactions the Andreev
fixed point is unstable and the normal one stable.  However, the reverse is true for attractive 
interactions.  This implies that a generic junction Hamiltonian
(without fine-tuned junction parameters) 
will renormalize to the normal fixed point for repulsive interactions but to the Andreev one for
attractive interactions. An exact mapping of our tight-binding model to
the Hubbard model with
a transverse magnetic field is used to help understand this behavior. 
We calculate the critical exponents, which are different at these
two different fixed points.   
\end{abstract}

\vspace{1.0cm}
\section{introduction}
One of the fascinating consequences of superconductivity is the
phenomenon of Andreev scattering at a normal metal/superconductor (NS)
junction.  This corresponds to an incoming electron from the normal
side being reflected back as a hole, thereby producing an additional
Cooper pair in the superconductor condensate.  Both normal and Andreev
reflections are expected to occur in a realistic NS junction, in
addition to quasiparticle transmission into the superconductor.  If
the gap is large enough, the latter's propagation is suppressed and
squared reflection amplitudes add up to one through probability
current conservation.  Building up on Andreev reflection, one arrives
at the related phenomenon of the Josephson current, in which the
normal region of an
SNS junctions carries a supercurrent driven by the gap phase
difference between left and right superconductors.

While the original work in this field\cite{AndreevZETF,DeGennesBook,DemersPRB4,BlonderPRB25}
treated the normal metal within the Fermi liquid framework ({\it i.e.}
essentially ignored interactions), the
effect of interactions for the case of a one-dimensional metal between
two superconductors has been treated 
recently by several groups\cite{MaslovPRB53,Fazio1,Takane,Fazio2}
 using bosonization and renormalization group methods.  The methods and conclusions 
of these works are closely related to previous work on tunnelling
through a single impurity in a quantum wire.\cite{Kane}

We have chosen to re-examine this subject in both non-interacting and
interacting cases because we feel 
that previous treatments have missed some interesting physics.  In particular, most of the
standard work on the non-interacting case has essentially ignored band
structure effects, using
a free electron model with a pairing potential that varies abruptly
accross the junction, together with
a scattering potential at the interface.  A more recent paper\cite{Deutscher}
 considers the case where the
Fermi velocity is different on the S and N side. The conclusion of
this work is that, at the Fermi energy, 
there is perfect Andreev reflection (and therefore 0 normal reflection) when the scattering potential
and velocity mismatch are absent.  Both of these effects 
serve to increase the normal scattering
amplitude in an additive way.  We arrive at qualitatively different conclusions by explicitly
including band structure in the form of an exactly solveable
 one-dimensional tight-binding model with Hamiltonian:
\begin{equation}
H-\mu N = \sum_{j} \left[ (-t_{j,j+1} \psi_{j \sigma}^{\dagger} \psi_{j+1 \sigma} + \Delta_j 
\psi_{j \uparrow}^{\dagger} \psi_{j \downarrow}^{\dagger} + h.c.) + 
 (V_j - \mu) \psi_{j \sigma}^{\dagger} \psi_{j \sigma} \right] \label{ifH}.
\end{equation}
The chemical potential, $\mu$, is assumed to lie within the band on the normal
side $|\mu |<2t$
 and $h.c.$ stands for Hermitean conjugate.
Here the interface is chosen to lie between sites 0 and 1 with the superconductor
on the negative $x$-axis and the normal metal on the positive $x$ axis so that:
\begin{eqnarray}
t_{j,j+1} =  \left\{ \begin{array}{ll} t & j>0 \\ t'' & j=0 \\ t' & j <0
\end{array} \right., \nonumber \\
V_j = V \delta_{j1}, \hspace{1.0cm}\Delta_j = 
\left\{ \begin{array}{ll} 
\Delta & j\leq 0 \\ 0 & j >0 
\end{array} \right. . \label{parameters}
\end{eqnarray}
$\Delta_j$ represents the pairing interaction
which exists on the superconducting side ($j\leq 0$) only.  For simplicity we consider
both normal and superconducting sides to be one-dimensional, but see below.
The notion of a ``perfect junction'' becomes less clear in such a model.  In the
particle-hole symmetric case, $\mu =V=0$, we find that there is always some
normal scattering at the Fermi energy 
unless the interface tunnelling parameter, $t''$, is fine-tuned
to a particular value.  In the limit $|\Delta |<<t,t'$, this particular value becomes
\begin{equation}
t''=\sqrt{tt'}.\end{equation}
For general values of the chemical potential we find that both $t''$ and the normal
scattering intensity, $V$, must be fine-tuned in order to achieve perfect Andreev
reflection.  We emphasise that these conclusions seem to be different than
the previous ones obtained without explicit consideration of band effects.  For example,
in the particle-hole symmetric case it is possible to get perfect Andreev reflection
even with Fermi velocity mismatch ($t\neq t'$) provided that $t''$ is adjusted to the right value. 
It is worth emphasizing that the case of infinite interface tunnelling:
$t''>>t,t'$ {\it does not} correspond to perfect Andreev reflection as one
might naively suppose, but instead
to zero Andreev reflection.  The physical reason is that, in this limit, two electrons
get trapped at the interface on sites 0 and 1, effectively
decoupling all sites with $j<0$ from all sites with $j>1$.  In this limit the normal
side does not ``feel'' the pairing and hence exhibits no Andreev reflection.

We are not aware of any previous explicit calculation of the Josephson current for such an interface 
model which we perform here. 
We consider two, possibly different, interfaces separated by a distance of $l$ lattice sites with the
pairing potentials having a phase difference $\chi$.  The (zero temperature)
 Josephson current is defined from
the derivative of the groundstate energy with respect to this phase difference:
\begin{eqnarray}
I(\chi) = 2e \frac{d}{d \chi} E_0.
\end{eqnarray}
While the groundstate energy is obtained by summing over all states below the
Fermi surface, we show explicitly that its derivative with respect to $\chi$
only depends on quantities defined at the Fermi surface, being insensitive to
the details of the band structure.  
When the junction parameters are fine-tuned to give perfect Andreev reflection
we find that the Josephson current is a sawtooth function of $\chi$:
\begin{eqnarray}
I(\chi) \to \frac{e v_f}{\pi l} \chi, \hspace{1.0cm} (\hbox{mod}\  2\pi), 
\end{eqnarray}
with steps of size $2ev_f/l$ occurring at $\chi = (2n+1)\pi$.
For any other choice of junction parameters the Josephson current is
a smooth function of $\chi$.  For example, in the limit of a weak junction, $t''<<t,t'$,
we find:
\begin{equation}
I(\chi )\propto (t'')^4\sin \chi .\end{equation}
While these formulas are well-known results, our method in fact allows
us to calculate exactly the zero-temperature Josephson current in the
general case of arbitrary amounts of normal versus Andreev reflection,
independently at each boundary.  The full crossover between the above
two limits is thus described.

One approach which we shall make much use of in this paper is that of
an effective field theory, whereby the superconducting side is
replaced by a particular boundary contribution on the normal side.
There are several reasons why it is convenient to ``integrate out'' the electrons on 
the superconducting side of the junction in such a way as to obtain an
effective Hamiltonian for the electrons
on the normal side.  This is a very natural thing to do considering
the fact that the superconducting electrons
have a gap in their spectrum: if we consider physics at energy scales small compared
to the gap we expect to obtain a simple effective action without any retarded interactions.  
(This may break down for non s-wave pairing where the gap vanishes in certain directions; we
do not consider that case here.)  The resulting effective Hamiltonian (for a single junction) is:
\begin{eqnarray}
H -\mu N =-t\sum_{j\geq 1}\left[(\psi_{j \sigma}^{\dagger} \psi_{j+1
\sigma}+  h.c.) -\mu \psi^\dagger_{j\sigma}
\psi_{j\sigma} \right]+\left[ \Delta_B
\psi^\dagger_{1,\uparrow}\psi^\dagger_{1,\downarrow}+h.c.\right]+V_B\psi^\dagger_{1\sigma}
\psi_{1\sigma}.\label{Heff}
\end{eqnarray}
The effective boundary pairing interaction, $\Delta_B$ and effective boundary scattering potential, $V_B$,
depend on all the parameters of the superconductor and the junction, $t'$, $\Delta$, $t''$
, $V$.  Beginning
from our interface model of Eq. (\ref{ifH}) we determine explicitly the parameters $\Delta_B$ and
$V_B$ of the boundary model and check that low energy properties are faithfully reproduced.  
Of course, we again find with the boundary pairing model that perfect Andreev reflection
only occurs if the boundary parameters are fine-tuned.  In particular, for the
particle-hole symmetric case, $\mu =V_B=0$, the condition is: 
\begin{equation}|\Delta_B|=t.\end{equation}
Note that is is {\it not} $\Delta_B\to \infty$ as one might naively suppose. 

One advantage of the boundary model is that it should arise from much more general, and more realistic,
interface models. 
 While the simple form of Eq. (\ref{ifH}), quadratic in fermion operators, is 
the result of a mean field approximation to a more realistic model with pairing or electron-phonon
interactions on the superconducting side,
 we expect that the effective Hamiltonian of Eq. (\ref{Heff}) will still be valid
at energies small compared to the gap when the interactions on the
superconducting side are treated more accurately.  Furthermore, it is more or less obvious that the same
effective Hamiltonian arises when the one-dimensional normal metal is coupled to a three-dimensional
superconductor.  This is an important generalization since superconductivity is not believed to
occur in a strictly one-dimensional system.  

More generally we wish to add interactions to our Hamiltonian on the normal side which
we assume to be one-dimensional.  A simple choice would be an onsite (Hubbard) interaction
for all sites $j\geq 1$:
\begin{eqnarray}
H&\to& H + H_{int}\nonumber \\
H_{int}&=& {U\over 2}\sum_{j\geq 1}(n_j-1)^2,\label{Hint}\end{eqnarray}
where $n_j$ is the total electron number operator at site $j$.  We could 
also consider
longer range density-density interactions.  We will be interested in the
case of both repulsive and attractive bulk interactions; the latter may
arise in a low energy effective theory from phonon exchange.  The negative $U$
Hubbard model has a gap for spin excitations which can be eliminated by
considering longer range interactions.  We will discuss both cases
with zero and non-zero spin gap.
  These interactions can be treated essentially exactly, at low energies, using bosonization,
renormalization group and conformal field theory techniques.  It is a major
purpose of this paper to discuss how to generalize these techniques to the interface
model.  We argue that this is best done by integrating out the superconducting
electrons to obtain the boundary model of Eq. (\ref{Heff}) with the bulk interactions,
$H_{int}$ added.  (More generally, we might also obtain additional interactions
at the boundary.  These can be treated in the same framework.)  Indeed, this approach
seems to be more or less forced upon us by the renormalization group philosophy of
integrating out high energy modes to obtain an effective low energy Hamiltonian.  
Having performed the initial step of integrating out the gapped degrees of freedom
on the superconducting side we may then proceed to analyse the effective
Hamiltonian with bulk
and boundary terms  using general methods developed to deal with quantum
impurity problems.\cite{Affleck1}  Of course,  our results will only be valid at energies
$E<<\Delta$.  

The boundary renormalization group approach leads to the conclusion that the
boundary interactions will renormalize to a fixed point corresponding to
a conformally invariant boundary condition.  It appears likely that there
are only two such boundary conditions that occur in this problem, in the
particle-hole symmetric case ($\mu = V = V_B=0$) corresponding
to a free boundary condition (b.c.), $\Delta_B=0$ which preserves electron number
and therefore has no Andreev reflection and to an ``Andreev boundary condition''
for which there is perfect Andreev reflection.  Which of these boundary
conditions is stable under renormalization group transformations depends on
the sign of the bulk interactions, $U$.  We find that the free b.c. is stable
for repulsive bulk interactions but the Andreev b.c. is stable for attractive
bulk interactions ($U<0$).  We calculate the various critical exponents associated
with these critical points.  It is important to realize that critical exponents
are characteristic of a particular fixed point and are different at the free
and Andreev fixed points.  Thus, for instance with repulsive
bulk interactions, if we started with bare interface
parameters that put the interface close to the Andreev fixed point then the
exponents characterizing the initial flow away from the Andreev fixed point
at high temperature are different than the exponents characterizing the flow
towards the free fixed point at low temperature.  However, it should be
emphasized that for a generic choice of bare interface parameters the Hamiltonian
will not be near the Andreev fixed point; this requires fine-tuning.  The
situation is similar in the non particle-hole symmetric case except we now
get lines of fixed points with either $0$ or perfect Andreev reflection.

The first work that we are aware of on Andreev scattering in Tomonaga-Luttinger 
liquids\cite{MaslovPRB53} attempted to apply boundary RG techniques without
explicitly intergrating out the superconducting side and without taking 
into account the effect of the b.c.'s on the exponents.  This led to incorrect
predictions for the exponent governing the Josephson current.  [\onlinecite{Fazio1}]
used a method closely related to ours but applied it to a different geometry: a closed
normal ring in contact with superconductors at two points.  Takane\cite{Takane} corrected 
some of the
earlier result in [\onlinecite{MaslovPRB53}] for the exponent
governing the Josephson current using methods essentially equivalent to ours 
but without explicitly invoking the concept of integrating out the superconducting
side.  We extend Takane's result by introducing the conceptually important and
very useful notion of integrating out the superconducting electrons thus making
clear the relationship between the interface problem, other quantum impurity problems
and boundary conformal field theory.  This facilitates a more general discussion of
the universal critical behaviour.  In particular we discuss the behaviour of
the Josephson current in the vicinity of the Andreev fixed point, obtaining quite
different results than those in [\onlinecite{MaslovPRB53}], and discussing
how the functional dependence of the current on the superconducting phase
difference crosses over between sawtooth and smooth forms.  

Given
the difficulty of achieving perfect Andreev scattering in the non-interacting
case our conclusion is quite remarkable that, with attractive bulk interactions,
a generic interface will renormalize to perfect Andreev scattering as $T\to 0$.  
It must be admitted that this conclusion 
is based on an unproven but widely made assumption
about RG flows and fixed points in the boundary sine-Gordon model which arises
here after bosonization.  In order to make some of our rather unintuitive
results seem more plausible we discuss an exact mapping of our boundary pairing
model (in the particle-hole symmetric case) into a Hubbard model with a
bulk magnetic field and a transverse boundary magnetic field.  In the transformed
model the Andreev boundary condition corresponds to one in which the boundary 
electron has a frozen transverse spin polarization.  

In the next section we give the solution of the interface and boundary models
and discuss their equivalence, in the case of zero bulk interactions.  Most of
the details of the equivalence are relegated to an appendix.  
In Sec. III we consider an $S_1NS_2$ system and calculate the Josephson current.
 In Sec. IV we include bulk interactions in
the boundary model and determine phase diagrams and critical exponents.  In Sec. V
we discuss the exact mapping onto the  Hubbard model with bulk and boundary
magnetic fields. 

\section{The lattice interface and boundary models}
In this section, we consider various lattice models of normal metal-superconductor
contacts in the case of zero bulk interactions.  
The generic Hamiltonian for all of these will be (\ref{ifH}), with various
geometries, sets of hopping strenths, and  potentials specified along the way.
The calculation procedure is similar to the usual one used in dealing with continuum
models:  the wavefunctions are found for each sector, and the matching conditions at
the contacts yield consistency equations from which the various reflection/transmission
coefficients are obtained.  

We start by performing the Bogoliubov-de Gennes transformation
\begin{eqnarray}
\psi_{j \sigma} = \sum_{\alpha} [u_{\alpha j} \gamma_{\alpha \sigma} - \sigma
v^*_{\alpha j} \gamma^{\dagger}_{\alpha -\sigma} ], 
\end{eqnarray}
where the quasiparticle operators satisfy $\{ \gamma_{\alpha \sigma},
\gamma^{\dagger}_{\alpha' \sigma'} \} =  
\delta_{\alpha \alpha'} \delta_{\sigma \sigma'}$ and $\alpha$ is a (real) 
quasimomentum 
index.  We obtain \cite{DeGennesBook}, by requiring the Hamiltonian to be
diagonal $(H = E_0 +
\sum_{\alpha} \epsilon_{\alpha} \gamma_{\alpha \sigma}^{\dagger} \gamma_{\alpha
\sigma} )$ the following lattice Bogoliubov-de Gennes equations:
\begin{eqnarray}
\epsilon_{\alpha} u_{\alpha j} &=& -t_{jj-1} u_{\alpha j-1} -t_{jj+1} u_{\alpha j+1} 
+(V_j-\mu) u_{\alpha j} + \Delta_{j} v_{\alpha j}
\nonumber\\
\epsilon_{\alpha} v_{\alpha j} &=& t_{jj-1} v_{\alpha j-1} +t_{jj+1} v_{\alpha j+1} 
-(V_j-\mu) v_{\alpha j} + \Delta^*_{j} u_{\alpha j}.
\label{latticeBDG}
\end{eqnarray}
The solutions of these equations in two particular geometries are presented below.
The emphasis is put on the calculation of the Andreev reflection coefficient and
on the low-energy properties of the system.

\subsection{The lattice interface model}
\begin{figure}
\begin{center}
\epsfig{file=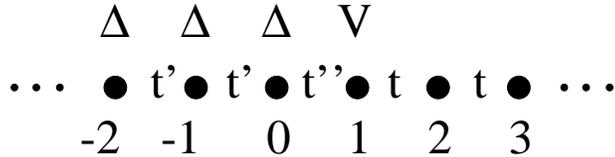, height=2cm}
\end{center}
\caption{The interface model.  Dots represent lattice sites, labeled
by integers.  On the left is the superconductor, with gap $\Delta$ and
bandwidth $t'$.  It is coupled to a normal metal with bandwidth $t$
through a junction of strength $t''$, with a local potential $V$ to
tune the normal scattering rate at the contact.}
\label{fig:interface}
\end{figure}

As a simple toy model for a NS interface, we consider the geometry depicted in
figure \ref{fig:interface}.  This tight-binding model is one of free electrons
in which a nonvanishing superconducting order parameter has been induced by an
unspecified mechanism on the left-hand sites of the lattice only.  Although
it is a straightforward lattice version of the ubiquitous continuum
one-dimensional models used in [\onlinecite{BlonderPRB25}] and
numerous subsequent work, 
let us underline that our model includes some additional features, namely 
different bandwidths $t, t'$ on the normal and superconducting sides, and
an arbitrary coupling $t''$ together with a local scattering potential
$V$ at the interface.  In other words, our lattice interface model is
defined by the BdG equations (\ref{latticeBDG}) with the choice of parameters
in (\ref{parameters}).

The calculation of the normal and Andreev reflection coefficients in the
lattice interface model is nothing but an exercise in elementary quantum
mechanics.  Namely, it is performed by choosing an appropriate ansatz for
the wavefunctions, after which solving for the matching conditions of the
eigenstates at the interface yields all the desired information.

Let us implement this procedure by taking a simple travelling wave
solution to the Bogoliubov-de Gennes equations (\ref{latticeBDG}):
\begin{eqnarray}
u_{\alpha j} = \left\{ \begin{array}{ll} e^{-i\alpha j} + R_N e^{i\alpha j} & j \geq 1 \\
T_N e^{-i\beta j} + (-1)^j T_N'e^{-i\delta j}& j \leq 0 \end{array} \right. \hspace{1.0cm}
v_{\alpha j} = \left\{ \begin{array}{ll} (-1)^j R_A e^{i\gamma j}& j \geq 1 \\
T_A e^{-i\beta j} + (-1)^j T_A' e^{-i\delta j} & j \leq 0 \end{array} \right. .
\label{solutions}
\end{eqnarray}
In the above equation, the quasimomenta $\beta$ and $\delta$ will turn out to be
complex for the region of parameters we will be concentrating on, namely at 
energies below the superconducting gap.  This means that the wavefunction amplitudes
$u_{\alpha j}$ and $v_{\alpha j}$ are exponentially decaying on the left-hand side.

The rest of the procedure is then a simple matter of 
substituting (\ref{solutions}) in (\ref{latticeBDG})
and carrying out the necessary algebra.  The energy is 
\begin{eqnarray}
\epsilon_{\alpha}=
-2t \cos \alpha -\mu,
\end{eqnarray}
while the other parameters in (\ref{solutions}) are given by
\begin{eqnarray}
\cos \gamma = \cos{\alpha} + \mu/t, \hspace{1.0cm}
2t'\cos{\beta} &=& \sqrt{(2t\cos{\alpha} + \mu)^2 - |\Delta|^2} -\mu, \nonumber \\
2t'\cos{\delta} &=& \sqrt{(2t\cos{\alpha} + \mu)^2 - |\Delta|^2} +\mu.
\end{eqnarray}

Although all reflection and transmission coefficients can be calculated explicitly, 
a simpler form for these expressions is obtained if we are interested primarily
in the scattering of particles whose energy is very small compared to the 
superconducting energy gap.  The $\epsilon \rightarrow 0$ limit gives
\begin{eqnarray}
&\cos{\alpha} =-\cos{\gamma} = -\frac{\mu}{2t},
&\sin{\alpha} = \sin{\gamma} = \sqrt{1-\mu^2/{4t^2}}, \nonumber\\
&\cos{\beta} = -(\cos{\delta})^* = -i |\Delta|/{2t'} -\mu/{2t'}, \hspace{1.0cm}
&\sin{\beta} = (\sin{\delta})^* = \sqrt{1-(i|\Delta| +\mu)^2/{4{t'}^2}},
\end{eqnarray}
and allows us to write the following expression for the Andreev reflection 
coefficient:
\begin{eqnarray}
R_A |_{\epsilon = 0} = i e^{-i \chi} \frac{t {t''}^2}{t'} \frac{\sqrt{1-\mu^2/4t^2} 
\left[ -\frac{|\Delta|}{t'} + \sin \beta + \sin \delta \right]}{t^2-\mu^2/4 + 
(V -\mu/2 -\frac{{t''}^2}{t'} e^{i\beta})(V-\mu/2 + \frac{{t''}^2}{t'} e^{i \delta})}.
\label{RAEOI}
\end{eqnarray}
In the above, $\chi$ is the phase of the order parameter, $\Delta = |\Delta| e^{i\chi}$.
Note that the reflection coefficients moreover obey the sum rule $|R_A|^2 + |R_N|^2 = 1$ 
at half-filling, in 
view of the conservation of the current $t_{j j+1} [u^*_{\alpha j+1} u_{\alpha j} -
v^*_{\alpha j+1} v_{\alpha j} - h.c.]$.

This low-energy Andreev reflection coefficient, as a function of the
bandwidths $t, t'$ and of the strength of the
pairing $\Delta$, as well as of the tunneling strength $t''$ and local scattering
potential $V$, obeys some simple but interesting properties.  Imagining for 
example a generic experimental setup in which variations in $t''$ and $V$ can 
be implemented, we can ask how $R_A|_{\epsilon = 0}$ behaves.  A simple
variation yields that the optimal amplitude is achieved for
\begin{eqnarray}
V_{max} = \frac{\mu}{2} + \frac{{t''}^2}{2t'} (e^{i\beta} +c.c.).
\end{eqnarray}
(Here $c.c.$ denotes complex conjugate.)
Subsequently tuning $t''$
yields a maximum at 
\begin{eqnarray}
{t''}_{max}^2 = 2 t t' \frac{\sqrt{1-\mu^2/4t^2}}{-\frac{|\Delta|}{t'} + (\sin
\beta + c.c.)}.
\end{eqnarray}
In the particle-hole symmetric case, this simplifies to 
\begin{eqnarray}
{t''}_{max}^2 = t t' \left[\sqrt{1+ \frac{|\Delta|^2}{4{t'}^2}} +
\frac{|\Delta|}{2t'} \right].
\end{eqnarray}
Tuning the parameters in such a way, we find perfect Andreev reflection, i.e.
\begin{eqnarray}
R_A {|_{\epsilon =0, V_{max}, t''_{max}}} = i e^{-i \chi}
\end{eqnarray}
In terms of left- and right-movers in the continuum limit of (\ref{solutions}), 
defined in Sec. IV, this corresponds to 
\begin{eqnarray}
\Psi_{R \uparrow}(0) = -i e^{i\chi} \Psi_{L \downarrow}^{\dagger}(0), \hspace{1.0cm} 
\Psi_{R \downarrow}(0) = i e^{i\chi} \Psi_{L \uparrow}^{\dagger}(0),
\label{and_bc}\end{eqnarray}
which are simply the perfect Andreev conditions usually imposed.

Thus, pure Andreev boundary conditions are obtainable in the lattice interface
model by simply tuning two parameters in the general non particle-hole symmetric
case.  For the particle-hole symmetric case, it is enough to tune one parameter.
 
\subsection{The lattice boundary model}
At energies below the superconducting gap, all wavefunctions incident on the
NS interface from the normal side are eventually reflected back through either
normal or Andreev reflection processes.  The vanishing of the transmission 
coefficients opens the door to the formulation of a different approach than
that adopted in the interface model, namely one in which we consider a system
with a boundary obtained by ``integrating out'' the superconducting side, 
yielding an effective pairing potential localized at the original contact.
In all generality, we will also consider a  potential present at the
boundary.

\begin{figure}
\begin{center}
\epsfig{file=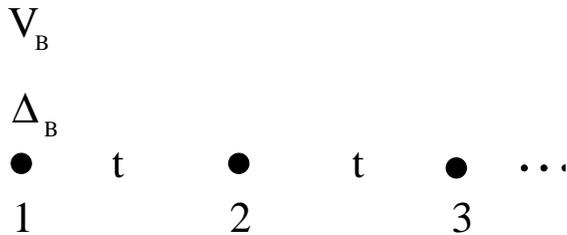, height=3.0cm}
\end{center}
\caption{The boundary model.   The superconductor has been replaced by
effective boundary potentials $\Delta_B$ and $V_B$.}
\label{fig:boundary}
\end{figure}

What we could call the lattice boundary model can be directly defined
through the BdG equations (\ref{latticeBDG}), 
but our system (illustrated in figure \ref{fig:boundary}) is now taken
to live on the
axis of positive integers, with a boundary at $j=0$ having on-site pairing $\Delta_B$ and 
potential $V_B$ on the first site.  
The hopping strength is taken to be $t$ between all sites.  For the
moment, we will perform all the relevant calculations from scratch,
deferring the explicit connection with the previous interface model
until later.

In the same way as for the interface model, we can calculate 
the normal and Andreev reflection 
coefficients by simple quantum mechanics.  Of primary interest is the
Andreev one, which is readily  
obtained by using the ansatz
\begin{eqnarray}
u_{\alpha j} = e^{-i \alpha j} + R_N e^{i \alpha j}, \hspace{0.5cm}
v_{\alpha j} = (-1)^j R_A e^{i \gamma j}
\end{eqnarray}
in the lattice BdG equations (\ref{latticeBDG}) in the geometry just described.
We find again $\epsilon_{\alpha} = -2t \cos \alpha -\mu$ with $\cos \gamma
= \cos \alpha + \mu/t$, together with
\begin{eqnarray}
R_A = 2i\frac{\Delta^*_B t \sin{\alpha}}{|\Delta_B|^2 + (t e^{-i\alpha} +V_B)
(-te^{-i\gamma} +V_B)}
\end{eqnarray}
which, at zero energy, becomes
\begin{eqnarray}
R_A |_{\epsilon =0} = 2i\frac{\Delta^*_B t \sqrt{1-\mu^2/4t^2}}{|\Delta_B|^2 + 
V_B(V_B-\mu) + t^2}.
\label{RAEOB}
\end{eqnarray}

\begin{figure}
\begin{center}
\epsfig{file=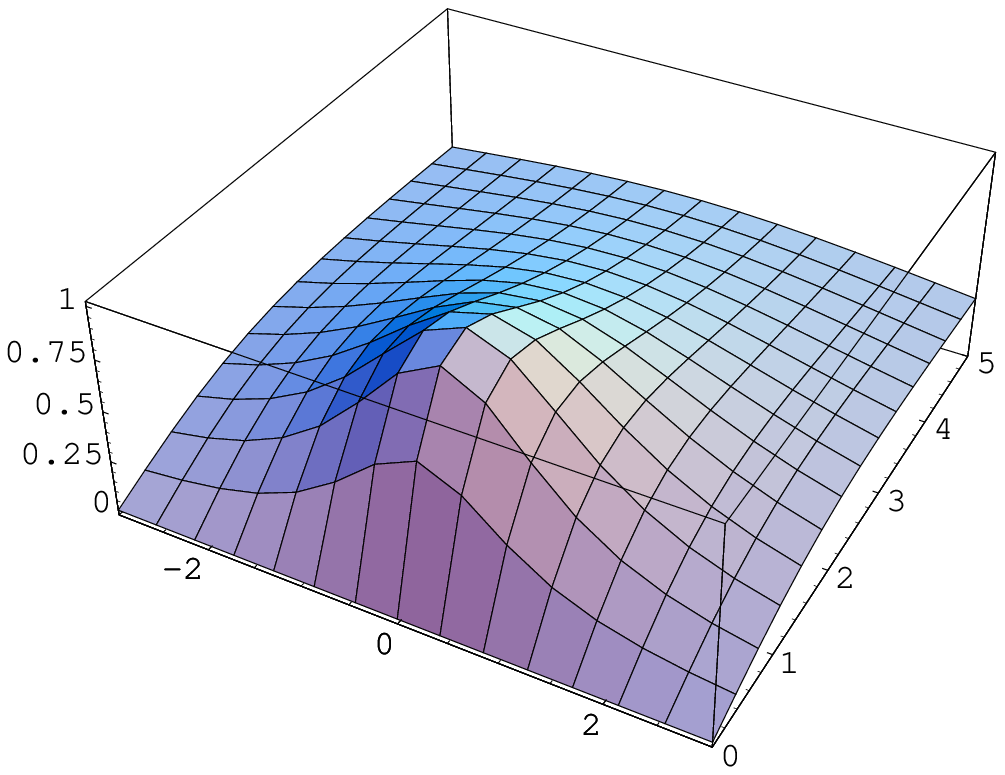, height=5cm}
\end{center}
\caption{$|R_A|$ for the boundary model at half-filling, 
plotted over the range $0 \leq |\Delta_B|/t \leq 5$ and $-3 \leq V_B/t \leq 3$.}
\label{fig:MODRA1}
\begin{center}
\epsfig{file=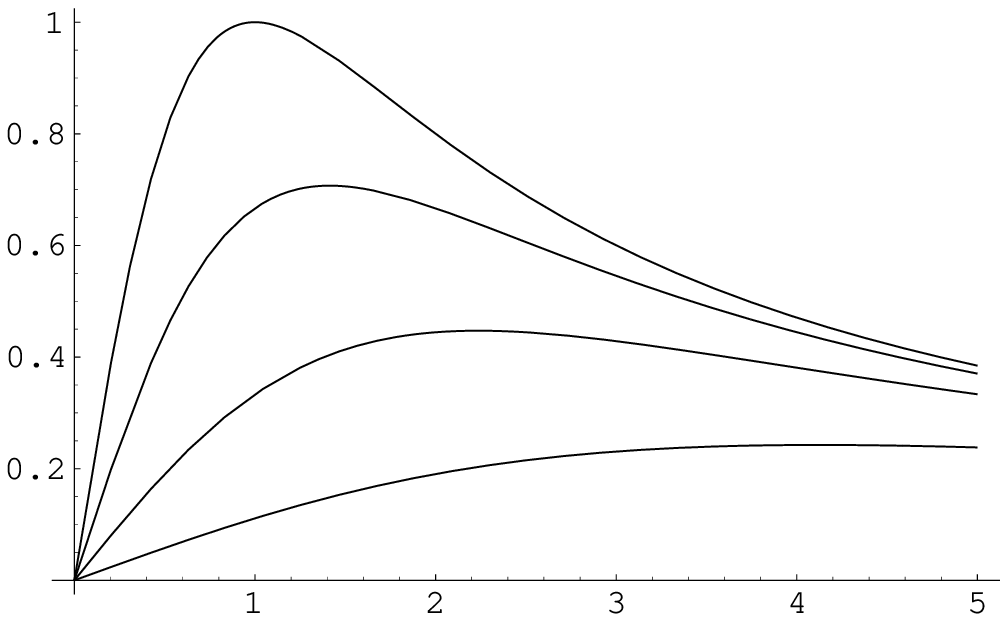, height=5cm}
\end{center}
\caption{$|R_A|$ plotted against $|\Delta_B|/t$ for $V_B/t = 0, 1, 2,$
and $4$.} 
\label{fig:MODRA3}
\end{figure}

The modulus of this coefficient is plotted in figure \ref{fig:MODRA1} at half-filling, as
a function of the boundary pairing $\Delta_B$ and boundary scattering potential
$V_B$.  The maximal amplitude has unit modulus, and such a maximum occurs for
any choice of filling in view of a similar optimization property to the one in
the interface model.  Namely, tuning $V_B$ and $\Delta_B$ yields a maximum amplitude at
\begin{eqnarray}
{V_B}_{max} = \mu/2, \hspace{1.0cm} {|\Delta_B|}_{max} = t \sqrt{1-\mu^2/4t^2},
\end{eqnarray}
and this once again produces the perfect Andreev condition
\begin{eqnarray}
{R_A}|_{\epsilon =0, {V_B}_{max}, {\Delta_B}_{max}} = i e^{-i \chi}.
\end{eqnarray}

Figure (\ref{fig:MODRA3}) shown the influence of a progressively stronger boundary 
scattering potential $V_B$ on the Andreev reflection coefficient at half-filling,
plotted against the boundary pairing strength.

The equivalence of formulas (\ref{RAEOI}) and (\ref{RAEOB}) for the effective
low-energy Andreev reflection coefficients comes from the fact that, as
mentioned before, the lattice boundary model is obtainable by
integrating out the gapped side 
of the interface model.  This procedure is outlined in Appendix A, in which the explicit
relationship between interface and boundary pairings and potentials is
derived.  At low energies, this correspondence reads
\begin{eqnarray}
\Delta_B = \frac{{t''}^2}{2t'} \left[ -\frac{|\Delta|}{t'} + \sin \beta 
+ \sin \delta \right], \hspace{1.0cm} 
V_B = V + \frac{{t''}^2}{2t'} \left[ \frac{\mu}{t'} - i\sin \beta 
+i \sin \delta \right].
\label{correspondence}
\end{eqnarray}
In the particle-hole symmetric case, this means that
\begin{eqnarray}
\Delta_B = e^{i\chi} \frac{{t''}^2}{t'} \left[\sqrt{1+|\Delta|^2/4{t'}^2} -|\Delta|/2{t'}
\right].
\end{eqnarray}

The interesting aspect of this formula comes from its behaviour in various limits.  
Contrary to simple intuition, a large bulk pairing $|\Delta| >> t'$ on the superconducting
side induces a ${\it small}$ boundary pairing $\Delta_B$, according to the limit
\begin{eqnarray}
\Delta_B  @>>{|\Delta| \gg t'}> e^{i\chi} \frac{{t''}^2}{|\Delta|},
\end{eqnarray}
whereas a small $|\Delta| << t'$ yields a boundary pairing whose amplitude
 depends on the
hopping parameters exclusively:
\begin{eqnarray}
\Delta_B @>>{|\Delta| \ll t'}> e^{i \chi} \frac{{t''}^2}{t'}.
\end{eqnarray}
It is important to note that these formulas hold for $\epsilon \ll |\Delta|$, so
one should not be surprised that a small bulk pairing still produces a significant
boundary pairing, with a finite amount of Andreev reflection.  The limits of zero
energy and zero pairing do not commute.
 
The above formulas allow one to move freely between interface and boundary formulations
of the NS problem, as long as the low-energy sector of the theory is considered.  They
will be used in the next section, which is devoted to the calculation of the Josephson
current in a SNS junction.

\section{Josephson current}

From the considerations of the earlier sections, we see that the problem of an 
$S_1 N S_2$ superconducting junction can be investigated within a double 
boundary framework, provided we are interested only in energies much smaller
than the gaps on either side.  If we imagine integrating out both the left and
right superconductors, we obtain a lattice model with two boundary pairings, 
which we dub the boundary junction model.  Namely, we take this to be the
model of free electrons in the bulk,
with pairings $\Delta_R$ and $\Delta_L$ at the right and left ends.  It is
important to realize that $\Delta_{R, L}$ are ${\it boundary}$ pairings, 
whose influence on the reflection coefficients has been explained in detail in
the previous section.  One should be careful not to confuse them with bulk
pairings, which have an altogether different effect (the two are related, of
course, by the relationship (\ref{correspondence})).  We thus 
define the system on sites $1, ..., l-1$ and take  
\begin{eqnarray}
\Delta_1 \equiv \Delta_L, \hspace{2.0cm} \Delta_{l-1} \equiv \Delta_R e^{i\chi},
\end{eqnarray}
with $\Delta_{R, L} \in {\Bbb R}$ (that is, we have put all the superconducting
phase difference on the right pairing).

Solving the lattice BdG equations by using the ansatz
\begin{eqnarray}
u_{\alpha j} = A_{\alpha} \sin \alpha j + B_{\alpha} \cos \alpha j, \hspace{1.0cm} 
v_{\alpha j} = (-1)^j \left( C_{\alpha} \sin \alpha j + D_{\alpha} \cos \alpha j
\right) 
\end{eqnarray}
with $\epsilon = -2t \cos \alpha$ yields after a certain amount of algebra
the condition for the allowed 
quasiparticle momenta $\alpha$ (we have chosen $l$ to be odd).
We find (for convenience, we have set  $t=1$ in what follows)
\begin{eqnarray}
0 = \sin^2 \alpha l + 2 \Delta_R \Delta_L \cos \chi \sin^2 \alpha 
-(\Delta_R^2 + \Delta_L^2)  \sin^2 \alpha(l-1) + \Delta_R^2 \Delta_L^2 
\sin^2 \alpha (l-2).
\label{allowedalpha}
\end{eqnarray}
Using a generalization of the approach used in
[\onlinecite{ZagoskinJPA30}] to treat 
the problem of free electrons on a tight-binding chain with a boundary 
scattering potential, we can conveniently find the closed form solution.  
First of all, let us write the allowed momenta in terms of
energy-dependent phase shifts $\delta_{\pm}$ as
\begin{eqnarray}
\alpha_{n \pm} = \frac{\pi n}{l} + \frac{\delta_{n \pm}}{l},
\end{eqnarray}
where $\pm$ refers to the two independent sets of Andreev levels,
labeled by the integer $n$.
Substituting this into (\ref{allowedalpha}) yields
\begin{eqnarray}
f_1 \cos 2\delta + f_2 \sin 2\delta = f_3 \hspace{2.0cm} 
&f_1 &= 1-(\Delta_R^2 + \Delta_L^2) \cos 2\alpha + \Delta_R^2 \Delta_L^2 
 \cos 4\alpha, \nonumber \\
&f_2 &= -(\Delta_R^2 + \Delta_L^2) \sin 2\alpha + \Delta_R^2 \Delta_L^2 
 \sin 4\alpha, \nonumber \\
&f_3 &= (1-\Delta_R^2) (1-\Delta_L^2) + 4 \Delta_R \Delta_L \sin^2
\alpha \cos \chi. 
\end{eqnarray}
Noting that $f_3^2 \leq f_1^2 + f_2^2$ throughout the parameter space allows
us to write this as
\begin{eqnarray}
\cos (2 \delta - g) = \cos h, 
\end{eqnarray}
where the functions $g$ and $h$ are defined by
\begin{eqnarray}
\cos g = \frac{f_1}{\sqrt{f_1^2 + f_2^2}}, \hspace{1.0cm} 
\cos h = \frac{f_3}{\sqrt{f_1^2 + f_2^2}}
\end{eqnarray}
Studying carefully the various functions along paths in the parameter space
yields a consistent choice of branches for the inverse trigonometric functions.
This leads to the final answer for the phase shifts, which we write as
\begin{eqnarray}
\delta_{\pm} = (g \pm h)/2
\label{delta}
\end{eqnarray}
where 
\begin{eqnarray}
&g = \left\{ \begin{array}{cc} -2\pi - \tilde{g}~~ & \alpha < \tilde{\alpha} \\
\tilde{g} & \tilde{\alpha} < \alpha < \pi - \tilde{\alpha} \\
2\pi - \tilde{g} & \pi -\tilde{\alpha} < \alpha \end{array} \right. 
\hspace{2.0cm}
&\tilde{g} = sgn(\alpha - \pi/2) \arccos \frac{f_1}{\sqrt{f_1^2 + f_2^2}}, 
\nonumber \\
&\tilde{\alpha} = \frac{1}{2} \Re \left\{\arccos{\frac{1}{2}(\Delta_R^{-2} + 
\Delta_L^{-2})}\right\}, \hspace{2.0cm}
&h = \arccos \frac{f_3}{\sqrt{f_1^2 + f_2^2}}
\end{eqnarray}
and all $arccos$ functions have their image in $[0, \pi]$. The phase
shifts are analytic except at the bottom and top of the band at the
perfect Andreev points, where $g$ suffers a branch jump.  

Knowing the phase shifts allows us to compute all the energy levels, and 
understand their behaviour as a function of the boundary pairings and the
superconducting phase difference $\chi$.  One recovers the usual picture 
wherein the levels are separated by finite gaps on an energy versus phase
diagram, except at the perfect Andreev points, where the gaps vanish.

The reason why it is so convenient to solve for the allowed quasiparticle
momenta in terms of these phase shifts, is that this procedure allows us to
write the ground-state energy straightforwardly in a $1/l$ expansion.  Again,
the derivation is very similar to the one in [\onlinecite{ZagoskinJPA30}], and we
refer the reader there for the missing details.

The ground-state energy can be written as the sum of the individual energies
of the occupied Andreev levels.  Namely,
\begin{eqnarray}
E_0 = \sum_{n=1}^N \sum_{\pm} \epsilon[\alpha_{n \pm}].
\end{eqnarray}
In the limit of large $l$, we can use a Euler-MacLaurin formula to transform
this sum into an integral.  Subsequently expanding to order $1/l$, we get
\begin{eqnarray}
E_0 = 2l \int_0^{k_F} \frac{dk}{\pi} \epsilon(k) + \frac{1}{\pi} \int_{\epsilon_0}^
{\epsilon_f} d\epsilon (\delta_+(\epsilon) + \delta_-(\epsilon))
+\frac{\pi v_F}{l} \left[ \frac{1}{2} \left( \frac
{\delta_+(k_F)}{\pi}\right)^2 + \frac{1}{2} \left( \frac
{\delta_-(k_F)}{\pi}\right)^2 - \frac{1}{12} \right] 
\end{eqnarray}
where $k_F = \pi(N+1/2)/l, \epsilon_F = \epsilon(k_F), \epsilon_0 = \epsilon(0) 
= -2, v_F = \epsilon'(k_F)$.
The crucial thing to notice here is that the $1/l$ terms are functions of data
exclusively at the Fermi surface.  While the $\frac{1}{12}$ term is well-known
to correspond to the finite-size contribution from open boundary 
conditions for a conformal field theory with central charge $c=2$ like
the present one (each spinful chiral fermion carries
a unit conformal charge), the other terms depending on the squares of the
phase shifts at $k_F$ give the change in $E_0$ coming from the effect of the
boundary pairings.  The $O(l^0)$ term, given by Fumi's theorem, depends however
on $\delta$ across the whole filled part of the band.

This expression for the ground-state energy allows us to write the Josephson
current at zero temperature in the limit of large $l$, in the presence of
arbitrary boundary pairings, i.e. with an arbitrary amount of normal versus
Andreev reflection on either edge.  Namely, the Josephson current is given by
\begin{eqnarray}
I(\chi) = 2e \frac{d}{d \chi} E_0.
\end{eqnarray}
Upon calculating this derivative, one easily sees that Fumi's theorem $O(l^0)$
term does not contribute to $I(\chi)$, since the sum of the phase shifts is
independent of $\chi$ for any energy (in (\ref{delta}), only $h$ depends on $\chi$).
Thus, the Josephson current is controlled
exclusively by parameters at the Fermi surface, and is given by the general
expression
\begin{eqnarray}
I (\chi) = \frac{e v_F^2}{\pi l} \Delta_R \Delta_L \frac{\sin \chi \arccos \left[
\frac{\tilde{g}(\cos \chi)}{(\tilde{g}^2(1) + \tilde{\Delta}^2v_F^2)^{1/2}}\right]}
{\sqrt{4\Delta_R \Delta_L 
\sin^2 \frac{\chi}{2} \tilde{g}(\cos^2 \frac{\chi}{2}) + \tilde{\Delta}^2}}
\label{current}
\end{eqnarray}
where we have defined 
\begin{eqnarray}
\tilde{g}(y) \equiv (1-\Delta_R^2)(1-\Delta_L^2) + \Delta_R \Delta_L v_F^2 y, 
\hspace{2.0cm}
\tilde{\Delta} \equiv (\Delta_R -\Delta_L) (1+ \Delta_R \Delta_L).
\nonumber
\end{eqnarray}
This function reproduces the well-known behaviours in the limiting cases of
perfect or very weak Andreev reflection:  the fine-tuning for perfect
Andreev reflection on both sides of the junction corresponds to
setting $\Delta_R = \Delta_L =1$, which, when substituted in 
(\ref{current}), yields
\begin{eqnarray}
I(\chi) @>>{\Delta_R, \Delta_L \rightarrow 1}>
\frac{e v_F}{\pi l} \chi, \hspace{1.0cm} |\chi| < \pi
\label{IperfectAndreev}
\end{eqnarray}
(Ishii's sawtooth), while on the other hand, for small pairing, we recover the
$\sin \chi$ behaviour:
\begin{eqnarray}
I(\chi) @>>{\Delta_R, \Delta_L << 1}> \frac{e v_F^3}{\pi l} \Delta_R
\Delta_L \sin \chi.
\end{eqnarray}

\begin{figure}
\begin{center}
\epsfig{file=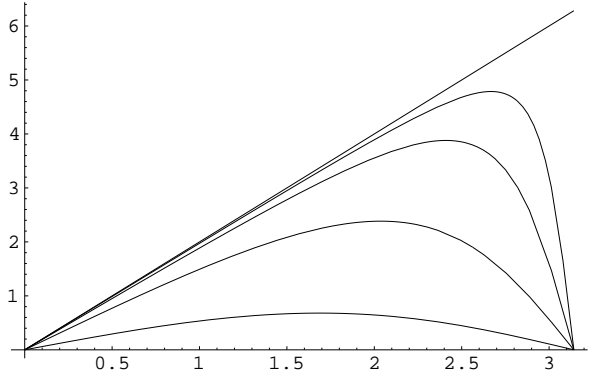, height=5cm}
\end{center}
\caption{Josephson current in the symmetric junction as a function of $\chi$, 
in units of $e/\pi l$.  The five plots are respectively for $|\Delta_B|/t = 1,
0.9, 0.8, 0.6 $ and $0.3$.}
\label{fig:IJOS1}
\end{figure}

It is instructive to plot (\ref{current}) in various regimes.  Taking symmetric
pairing $\Delta_R =\Delta_L =\Delta_B$ to start with, we can see how Ishii's
sawtooth is rounded off progressively to a $\sin \chi$ function as $\Delta_B$
is taken from 1 (perfect Andreev) to smaller and smaller values (i.e. for 
progressively more normal reflection at the contacts).  This is illustrated
in figure \ref{fig:IJOS1}.

It is important to note that the expression (\ref{current}) for the Josephson
current is valid for independent arbitrary values of the boundary pairings, and
thus covers the case of asymmetric junctions already studied for example in 
[\onlinecite{ChangPRB49}].  In this work, the shape of the
current-phase relationship 
was still the sawtooth function, with critical current depending on the 
asymmetry between the pairings.  The sawtooth result implies that effective
perfect Andreev conditions were imposed, and thus that no normal reflection
occurred at the contacts.  Our expression for the current in a long junction
thus covers a wider regime than the one in [\onlinecite{ChangPRB49}].
When plotting 
(\ref{current}) for various asymmetries, the graphs look very similar to
figure \ref{fig:IJOS1}.  

The formidable looking expression (\ref{current}) can be considerably simplified when
$\Delta_L\approx \Delta_R\approx 1$.  We first state the approximations
and then justify them afterwards.  We can approximate:
\begin{equation}
\arccos\left[ {\tilde g(\cos \chi )\over(\tilde g^2(1)+\tilde \Delta^2v_F^2)^{1/2}}\right] 
\approx \chi .\end{equation}
We may Taylor expand the $\chi$ dependence of the other factors near $|\chi |\approx \pi$
since they give essentially a constant elsewhere.  Thus:
\begin{equation}
\tilde g\left( \cos^2{\chi \over 2}\right) \approx
4(1-\Delta_R)(1-\Delta_L)+{v_F^2(\pi -|\chi |)^2\over 4}, 
\end{equation}
giving:
\begin{equation}
I(\chi )\approx {ev_F\over \pi l}\chi {\pi -|\chi |\over \sqrt{(\pi
-|\chi |)^2+{4 (2-\Delta_R-\Delta_L)^2 
\over v_F^2}}}, \ \  |\chi |<\pi .\label{Iapprox}\end{equation}
Note that the last factor vanishes at $|\chi |\to \pi$ but approaches
1 for $\pi -|\chi |>>|2-\Delta_R -\Delta_L|$.  
For small $|2-\Delta_R -\Delta_L|$ we find that the maximum current occurs at:
\begin{equation}\pi -|\chi_M |\approx \pi^{1/3}\left[
{2(2-\Delta_L-\Delta_R)\over v_F}\right]^{2/3} 
\end{equation}
and has a value:
\begin{equation}
I_c={ev_F\over l}\left\{ 1-{3\over
2}\left[{2(2-\Delta_L-\Delta_R)\over \pi v_F}\right]^{2/3}\right\}. 
\end{equation}
Now let us consider the justification for these approximations.  
First note that:
\begin{equation}
{\tilde g(\cos \chi )\over(\tilde g^2(1)+\tilde \Delta^2v_F^2)^{1/2}}
=[\cos \chi + O(\epsilon^2)][1+O(\epsilon^2)],
\end{equation}
where, for convenience, we have defined:
\begin{equation}
\epsilon \equiv (2-\Delta_L-\Delta_R)/v_F.\end{equation}
Thus, 
\begin{equation}
\arccos\left[ {\tilde g(\cos \chi )\over(\tilde g^2(1)+\tilde \Delta^2v_F^2)^{1/2}}\right] 
= \chi +O(\epsilon^2),\end{equation}
except near $|\chi |\approx \pi$ where $\arccos$ becomes singular, behaving as:
\begin{equation}
\arccos (-1+\delta ) \approx \pi -\sqrt{2\delta}.\end{equation}
Thus, near $|\chi |\approx \pi$ we may write:
\begin{equation}
\arccos\left[ {\tilde g(\cos \chi )\over(\tilde g^2(1)+\tilde \Delta^2v_F^2)^{1/2}}\right] 
\approx \pi -\sqrt{(\pi -|\chi |)^2+O(\epsilon^2)}.\end{equation}
Noting that, at the maximum $\chi_M$, $(\pi -|\chi |)\propto \epsilon^{2/3}$ we see that
in this range of $\chi$:
\begin{equation}
\arccos\left[ {\tilde g(\cos \chi )\over(\tilde g^2(1)+\tilde \Delta^2v_F^2)^{1/2}}\right] 
\approx \chi [1+ O(\epsilon^{2/3})].\end{equation}
Thus our simple approximation to $\arccos$ is everywhere valid.  The other
corrections from expanding $\sin \chi$ in the numerator in Eq. (\ref{current}) and
the expression in the denominator give multiplicative corrections of $O(\epsilon )$
or $O(\pi -|\chi |)^2$ which is $O(\epsilon^{4/3})$ near $\chi_M$.

In the case of perfect Andreev reflection at both boundaries, we can in fact
solve for the ground state energy (and thus the Josephson current) exactly for 
arbitrary junction length $l$ and filling.  Putting $\Delta_R =
\Delta_L = t (=1)$
in (\ref{allowedalpha}) directly yields after simple algebra the allowed 
quasiparticle momenta for the two sets of Andreev levels.  We find
\begin{eqnarray}
\alpha_{n \pm} = \frac{\pi n \pm \chi/2}{l-1}.
\end{eqnarray}
The ground state energy is then again the sum of the energies of all occupied
levels, which now becomes a simple geometric progression:
\begin{eqnarray}
E_0 = -2t \sum_{n=1}^N [\cos \alpha_+ + \cos \alpha_-] = -4t 
\frac{\sin \frac{\pi N}{2(l-1)} \cos \frac{\pi (N+1)}{2(l-1)}}{\sin \frac{\pi}{2(l-1)}}
\cos \frac{\chi}{2(l-1)}.\ \  (|\chi |<\pi )
\end{eqnarray}
The Josephson current is then, for arbitrary length $l$ and occupation number
$N$,
\begin{eqnarray}
I (\chi) = \frac{4 e t}{l-1} \frac{\sin \frac{\pi N}{2(l-1)} \cos 
\frac{\pi (N+1)}{2(l-1)}}{\sin \frac{\pi}{2(l-1)}} \sin \frac{\chi}{2(l-1)}.
\ \  (|\chi |<\pi )\end{eqnarray}
One can explicitly check that the large $l$ limit reproduces (\ref{IperfectAndreev}).

\section{Renormalization group analysis of the interacting case}
In a standard way the bulk Hamiltonian can be approximated in the continuum limit,
valid at low energies, by a quantum field theory, corresponding to a
Tomonaga-Luttinger liquid describing gapless charge and spin bosons.
  The first step is to write the lattice fermion
operators in terms of left and right moving continuum fermion operators:
\begin{equation}
\psi_{j\sigma}\approx  e^{-ik_Fx}\psi_{L\sigma}(x)+
e^{ik_Fx}\psi_{R\sigma}(x),\label{cont_lim}\end{equation}
where $k_F$ is the Fermi wave-vector and $\psi_{L,R}$ are assumed to vary slowly
on the scale of the lattice spacing (which is set to 1).
The resulting continuum Hamiltonian is then bosonized in terms of charge and
spin boson, $\phi_{c,s}$ with associated velocities, $v_{c,s}$ and compactification
radii $R_{c,s}$.  We will normally set these velocity parameters to 1.  The 
continuum fermion fields are written:
\begin{eqnarray}
\psi_{L\uparrow ,\downarrow}&\approx &\exp \left[ -i\left( {\phi_c\over 2R_c}
+\pi R_c\tilde \phi_c\pm {\phi_s\over 2R_s}\pm \pi R_s\tilde \phi_s\right) \right]
\nonumber \\
\psi_{R\uparrow ,\downarrow}&\approx &\exp \left[ i\left( {\phi_c\over 2R_c}
-\pi R_c\tilde \phi_c\pm {\phi_s\over 2R_s}\mp \pi R_s\tilde \phi_s\right) \right].
\end{eqnarray}
Here the bosons and dual bosons are written in terms of left and right moving
components as:
\begin{equation}
\phi (t,x)= \phi_L(t+x)+\phi_R(t-x),\ \ \tilde \phi (t,x)= \phi_L(t+x)-\phi_R(t-x).
\end{equation}
The Lagrangian has conventional normalization:
\begin{equation}
{\cal L}= (1/2)[\partial_{\mu}\phi_c\partial^{\mu} \phi_c]
+(1/2)[\partial_{\mu}\phi_s\partial^{\mu} \phi_s].
\end{equation}
Here we follow the conventions of [\onlinecite{Wong}].  
Unfortunately, various other bosonization conventions are frequently used.  In particular,
the compactification radii, $R_{c,s}$, which depend on the bulk interactions, are
often removed from the bosonization formulae by rescaling the bosons, resulting
in an unconventional normalization of the two terms in the Lagrangian.  The 
resulting normalization constants are sometimes called $g_{\rho,\sigma}$.  The relationship
between parameters $g_{\rho ,\sigma}$ used in [\onlinecite{Kane,Fazio1,Fazio2,Takane}], the
parameters $K_{\rho ,\sigma}$ used in [\onlinecite{MaslovPRB53}] and
our parameters is:
\begin{equation}
\pi R_{c,s}^2={1\over g_{\rho ,\sigma}}={1\over 2K_{\rho ,\sigma}}.
\end{equation}
In the case of $SU(2)$ symmetry, $R_s={1\over \sqrt{2\pi}}$.
For replusive bulk interactions $R_c>1/\sqrt{2\pi}$ and for attractive
bulk interactions $R_c<1/\sqrt{2\pi}$.  In the case of attractive interactions,
there may be a gap for spin excitations depending on the detailed form
of the bulk interactions.  This occurs, for example, for the attractive
Hubbard model.  The presence of a bulk spin gap makes very little difference
to our analysis.  Essentially, we may just drop $\phi_s$ from our
formulas.  We consider both cases below. 

Free boundary conditions, which occur for $\Delta_B=0$, correspond to:\cite{Eggert}
\begin{equation}
\psi_L(0)=e^{-i\theta}\psi_R(0),\end{equation}
where the phase $\theta$ depends on the boundary scattering potential $V_B$.
Note that these boundary conditions correspond to only normal reflection,
and thus zero Andreev reflection, so in this context it is appropriate to refer to them
as ``normal'' b.c.'s.  
In bosonized form these b.c.'s become:
\begin{equation}
\phi_c(0)=R_c\theta ,\ \  \phi_s(0)=0.\label{bcn}
\end{equation}
It is crucial to realise that these equations imply that $\phi_{Rc,s}$ may
be regarded as the analytic continuation of $\phi_{Lc,s}$ to the negative $x$-axis:
\begin{equation}
\phi_{cR}(x)=-\phi_{cL}(-x)+R_c\theta ,\ \  \phi_{sR}(x)=-\phi_{sL}(-x),\ \  (x>0).
\end{equation}
In particular, this implies 
\begin{equation}\tilde \phi_{c}(0)\to 2\phi_{Lc}(0)-R_c\theta .
\end{equation}
We now wish to consider the bosonized form of the boundary scattering potential,
$\propto V_B$ and boundary pairing interaction, $\propto \Delta_B$ in Eq. (\ref{Heff}).
The scattering potential is proportional to $\partial_x\phi_c (0)$.  This
has scaling dimension 1 and hence is marginal.  Note that boundary interactions
are relevant if their dimension is $d<1$ and irrelevant if $d>1$.  This is different
than for bulk interactions in (1+1) dimensions due to the fact that boundary
interactions are only integrated over time, not space.  Thus a boundary scattering
potential leads to a line of fixed points, characterized by a phase shift.  

On the
other hand, the boundary pairing interaction leads to the term:
\begin{equation}
\Delta_B^*\epsilon^{\alpha \beta}\psi_{L\alpha}\psi_{R\beta}+ h.c.\propto 
|\Delta_B|\sin[2\pi R_c\tilde \phi_{c}(0)+\chi ]\cos [\phi_s(0)/R_s]\to \Delta_B
\cos[4\pi R_c\phi_{Lc}(0)-2\pi R_c^2\theta +\chi ],\end{equation}
where the b.c. of Eq. (\ref{bcn}) was used in the last step and $\chi$ is the phase of $\Delta_B$.  
This operator has dimension $2\pi R_c^2$ and hence is irrelevant for repulsive bulk interactions
but relevant for attractive bulk interactions.  Thus we reach the important conclusion
that a weak boundary pairing interaction, $\Delta_B$,
 becomes progressively less important as $T\to 0$
in the case of repulsive bulk interactions.  This implies that the effective
coupling of the superconductor to the Luttinger liquid, $t''$, renormalizes to 0 since
$\Delta_B\propto (t'')^2$.  

In the case of attractive bulk interactions, $R_c<1/\sqrt{2\pi}$ the free boundary
condition is an unstable fixed point.  The ``obvious'' guess is that the Hamiltonian
renormalizes to a boundary fixed point corresponding to the boundary condition
\begin{equation}
\tilde \phi_c(0)=-(\chi +\pi /2)/2\pi R_c,\ \  \phi_s(0)=0.\end{equation}
Note that the boundary condition on the spin boson is unchanged.  This is surely
a reasonable assumption since the boundary interaction doesn't involve the
spin boson.  In fact this boundary condition is fixed by SU(2) symmetry.  
On the other hand, we are assuming that the effect of the relevant
boundary sine-Gordon interaction is to pin the dual charge boson, $\tilde \phi_c(0)$,
corresponding to a semi-classical analysis of the interaction at large $\Delta_B$.  
We note that the analogous assumption has been made in several other 
contexts.\cite{Kane,Eggert,Affleck2}  It is generally believed that only Dirichlet
and Neumann fixed points occur in the boundary sine-Gordon model for generic
compactification radius.  [We note that $\phi =$ constant corresponds to
a Dirichlet b.c. and $\tilde \phi =$ constant to a Neumann b.c. using the 
fact that $\partial \tilde \phi /\partial t=\partial \phi /\partial x$.]  In
order to shed more insight on this assumption, we discuss, in the next section,
a different boundary model which is equivalent to this one under an exact duality
transformation.  In the present context this boundary condition corresponds to
perfect Andreev reflection since it follows from Eq. (\ref{and_bc}).

The consistency of this assumption can be checked by considering the 
renormalization group stability of the Andreev b.c. Note that the
scaling dimension of boundary operators are different at this fixed point 
where we must use:
\begin{equation}
\phi_{cR}(x)=\phi_{cL}(-x)+(\chi +\pi /2)/2\pi R_c,\ \  \phi_{sR}(x)=-\phi_{sL}(-x),\ \  (x>0).
\label{bca}\end{equation}
In this case a further boundary pairing interaction is marginal, corresponding
to shifting the condensate phase, $\chi$.  The potentially relevant
interaction corresponds to normal scattering.  This corresponds to adding a
term
\begin{equation}
\delta H = V_N\psi^\dagger_{L\sigma}(0)\psi_{R\sigma}(0)+ h.c.
\label{V_N}\end{equation}
to the Hamiltonian obeying the Andreev b.c.  Using the Andreev boundary condition,
and letting $\theta$ be the phase of $V_N$, 
this term reduces to:
\begin{equation}
\delta H \propto -|V_N| \sin [\phi_c/R_c+\theta ] \propto 
-|V_N|\sin [2\phi_{cL}/R_c+(\chi +\pi /2)/2\pi R_c^2+\theta ],
\end{equation}
of dimension $1/(2\pi R_c^2)$.  This is irrelevant for attractive bulk interactions, but 
relevant in the repulsive case.  Thus we conclude that the Andreev b.c.
represents an attractive fixed point with attractive bulk interactions, so
that our assumption that a boundary pairing interaction leads to a flow to
the Andreev b.c. for $R_c<1/\sqrt{2\pi}$ is consistent.  On the other hand,
in the case of repulsive bulk interactions we expect an RG flow from
the Andreev fixed point to the normal fixed point.  For the non-interacting
case, both fixed points are marginal and no renormalization occurs.  There
is a line of fixed points connecting normal and Andreev fixed points along
which the ratio of normal to Andreev scattering varies continously.  This
behaviour is very analogous to the backscattering problem for a
single impurity in a quantum wire.\cite{Kane}

The behaviour of the Josephson current in the case of attractive bulk
interactions is especially interesting.  From Sec. 3 we see, that 
for the non-interacting case, the Josephson current is a sawtooth
function of amplitude $ev_F/l$ when the boundary terms are fine-tuned
to give perfect Andreev scattering at the Fermi surface.  Otherwise,
$I(\chi )$ is a smooth function.  As shown by Maslov et al.\cite{MaslovPRB53},
$I(\chi )$ is also a sawtooth function in the presence of bulk
interactions if the Andreev b.c. is applied to the bosonized theory,
with the amplitude replaced by $ev_F/(2\pi R_c^2l)$. The sawtooth
form is a universal property of the Andreev fixed point.  This
universality of the $O(1/l)$ term in the groundstate energy is a
familiar aspect of conformal field theory.  In cases where
the boundary parameters are not fine-tuned to the Andreev boundary
condition, but instead the Hamiltonian renormalizes to the Andreev
fixed point we expect to recover the same sawtooth form of
$I(\chi )$ in the limit $l\to \infty$.  However, the finite size
corrections will smooth out $I(\chi )$ since a finite $l$ cuts
off the RG flow of the junction parameter $\Delta_B$.  We 
expect that for large enough $l$ we may use the expression 
Eq. (\ref{Iapprox}) with $1-\Delta$ replaced by an effective
value at scale $l$.  We may identify:
\begin{equation}
1-\Delta_B\propto V_N,\end{equation}
where $V_N$ is the effective normal scattering interaction 
introduced in our discussion of the RG stability of the Andreev 
fixed point in Eq. (\ref{V_N}).  This identification is reasonable
since it can be checked, for the non-interacting case, that
the normal reflection amplitude vanishes linearly in
$1-|\Delta_B|$.  Thus we expect that:
\begin{equation}
1-|\Delta_{Beff}(l)|\propto {1\over l^{1/(2\pi R_c^2)-1}}.\end{equation}
Substituting this expression into Eq. (\ref{Iapprox})
gives an approximate expression for the Josephson current.  As $l$
increases, $I(\chi )$ becomes a more and more rapidly varying function near 
$|\chi |\approx \pi$.  The maximum of $I(\chi )$ occurs at:
\begin{equation}
|\chi_M|\propto \pi - {1\over l^{2[1/(2\pi R_c^2)-1]/3}}\end{equation}
and the critical current scales with $l$ as:
\begin{equation}
I_c\approx {ev_F\over 2\pi R_c^2l}\left[1-{\hbox{constant}
\over l^{2[1/(2\pi R_c^2)-1]/3}}\right]. \end{equation}
The smoothing of $I(\chi )$ for finite size junctions due to renormalization
effects is quite distinct from the finite size effects that occur
in the non-interacting case, discussed in Sec. III.  These are suppressed by
powers of $1/l^2$ and {\it do not} smooth out the sawtooth structure
of $I(\chi )$.  In particular the maximum remains at $|\chi_M|=\pi$.  

We note that our result for the behaviour of the Josephson current near the
Andreev fixed point is very different from that obtained in [\onlinecite{MaslovPRB53}]
although both treatments use bosonization and RG arguments.  The difference
arises in part because we take into account the Andreev b.c. in calculating
the RG scaling of the normal reflection amplitude and in part because we take 
into account the singular
dependence of the Josephson current on the normal reflection amplitude.

In the case of repulsive bulk interactions and almost perfectly fine-tuned
junction parameters a flow away from the Andreev b.c. occurs with
increasing junction length.  In this case the effective parameter
$1-|\Delta_B(l)|$ {\it increases} with increasing $l$ so that
the sawtooth singularity is smoothed out as $l$ {\it increases}.  

As mentioned above, in the case of attractive bulk interactions a spin gap
sometimes occurs, for example in the $U<0$ Hubbard model.  This has
essentially no effect on the boundary RG discussed above since the
spin boson didn't play any role.  Essentially the spin boson is assumed
to always obey the Dirichelt b.c., $\phi_s(0)=0$ throughout the
RG flow which only affects the b.c.'s on the charge boson.  In the case
where there is a spin gap,  $\phi_s(x)$ is pinned at all points in
space; this is completely compatible with the assumption about the b.c.

We find the flow to the Andreev b.c. in the attractive case especially
remarkable because, as explained in the previous section, in the non-interacting
case perfect Andreev scattering can only be achieved by fine-tuning parameters.
Thus, in the interacting case, the RG flow must ``find'' the special value of
the parameters at which the normal scattering vanishes.  

\section{duality transformation}
In an effort to make more plausible the conjectures about RG flows in the
previous section and in order to make contact with previous work on
quantum impurity problems we present in this section an exact duality
transformation from the lattice boundary pairing model with bulk interactions 
of the previous section to a lattice model with both bulk and boundary
magnetic fields.  This is related to the previously studied\cite{Affleck2} S=1/2 xxz 
chain with a transverse boundary field.  These latter models are perhaps
easier to understand intuitively because semi-classical approximations
hold to some extent.  Furthermore there is an instructive difference
between the Hubbard chain and pure spin chain corresponding to a sort
of breakdown of ``spin-charge separation'' for strong boundary fields.

We begin with the (semi-infinite) boundary pairing of Eq. (\ref{Heff}) with
the Hubbard interaction of Eq. (\ref{Hint}) added.  We then
apply the well-known duality transformation which changes the sign of
the Hubbard coupling constant, $U$, and interchanges charge and spin operators.
  This is essentially a particle-hole transformation {\it for spin up
electrons only}:
\begin{equation}
\psi_{j\uparrow}\to (-1)^j\psi_{j\uparrow}^\dagger,\ \ 
\psi_{j\downarrow}\to \psi_{j\downarrow}.\end{equation}
This maps the hopping term into itself and the Hubbard interaction into
$(-1)\times$ itself.  The chemical potential term is mapped into
a magnetic field in the $z$-direction:
\begin{equation}
\psi^\dagger_{j\alpha}\psi_{j\alpha}\to 1-\psi^\dagger_{j\alpha}
(\sigma^z)_{\alpha \beta}\psi_{j\beta}.
\end{equation}
Thus a non-zero chemical potential, corresponding to average particle number
$<n_j>\neq 1$ maps into a non-zero bulk magnetic field in the $z$-direction.
Note however, that the dual model has zero chemical potential so it remains
at half-filling.  Longer range density-density interactions map into 
$z$-$z$ magnetic exchange interactions.  
The boundary scattering term, $V_B$ maps into a modified boundary field
in the $z$-direction.
The boundary pairing interaction is mapped into:
\begin{equation}
H_B\to \Delta_B^* \psi_{1\uparrow}^\dagger \psi_{1\downarrow}+ h.c.\end{equation}
This corresponds to a boundary magnetic field lying in the xy plane, transverse
to the bulk field, of magnitude $2|\Delta_B|$ and direction determined by the
phase of $\Delta_B$.  

This dual model is especially easy to analyse in the case where $U<0$ so that
there is a spin gap in the Hubbard model.  The dual model, with $U>0$ and
half-filling has
a gap for charge excitations.  The remaining gapless spin excitations are
approximately described by the Heisenberg model with the appropriate
magnetic fields.  To make this correspondance more precise, when $|U|>>t$,
the correspondance holds for the lattice models with an effective Heisenberg
exchange interaction $t^2/U$.  For smaller $U$ the correspondance still
holds for the low energy degrees of freedom.  Even in situations where
the original spin excitations were not gapped so that the dual charge
excitations are not gapped, we might expect some sort of correspondance
with the Heisenberg model at low energies due to spin-charge separation.  

The xxz S=1/2 spin model with a transverse boundary field (but no bulk field)
was analysed in [\onlinecite{Affleck2}].  There it was shown that the
bosonized version is the boundary sine-Gordon model with a 
boundary interaction which is relevant along the entire bulk xxz
critical line and it was conjectured
that an RG flow to the Neumann b.c. occurs.  In the particular case
of the xx model this can be proven exactly using Ising model duality
transformations.\cite{Guinea}  
The semi-classical interpretation of the Neumann
b.c. in this case is that the boundary spin is polarized in the direction
of the boundary field.  This analysis can be easily extended to include
a bulk magnetic field in the z-direction.  As shown in [\onlinecite{Affleck2}],
the dimension of the transverse boundary field is $2\pi R^2$ where $R$ is
the compactification radius of the boson in the spin chain.  [To fix our
conventions, the transverse staggered correlation exponent is also $2\pi R^2$.]
This becomes marginal for the isotropic xxx model but is relevant along the
entire (zero field) xxz critical line.  
The dependence of the radius on a magnetic field applied to the Heisenberg
model has been calculated from the Bethe ansatz.\cite{Fledderjohan,Cabra}
  The effect is again to
decrease the radius, hence making the transverse boundary field relevant.  
Thus it is again natural to conjecture a flow to a fixed boundary spin
polarized in the direction of the boundary field.  Thus, {\it flow to the
Andreev b.c. in the boundary pairing model is dual to flow to a fixed 
spin b.c. in the boundary field model}.  

This analysis can be extended to more general models with longer
range bulk interactions.  In particular,
we may consider cases in which the spin excitations are not gapped in
the original model so that charge excitations are not gapped in the
dual model.  Again it seems plausible that even a weak transverse
boundary field produces a flow to a polarized spin boundary fixed point.  
However, we now encounter another interesting phenomenon.  If the
boundary field is too strong it suppresses this RG flow.  This can be
seen from the fact that, in the limit of a very strong transverse boundary
field, one electron gets trapped on the first site with probability 1
in a state with spin polarized along the transverse field direction.  
Since the hopping term adds or removes an electron from site 1 it produces
a high energy state, with energy of order the boundary field, $|\Delta_B|$.  
All such processes are suppressed for $|\Delta_B|>>t$ meaning that the
first site decouples from all the others which therefore obey a free b.c.  
Thus, in the finite $U$ model the spin-polarized Neumann b.c. should
not be thought of as occuring at infinite boundary field, but rather at 
a finite value.  On the other hand, in the Heisenberg model we may indeed
think of the spin polarized fixed point as occurring at infinite
boundary field since the magnetic exchange interaction isn't suppressed by
the strong field.  Thus we see that the limit $U\to \infty$ and
$|\Delta_B|\to \infty$ do not commute.  

The above observation provides another way of understanding the
perhaps surprising discovery in the previous sections that the Andreev
fixed point does not occur at $\infty$ boundary pairing strength but
rather at a fine-tuned finite value.  In this model at very strong
$\Delta_B$ we may think of a sort of Andreev boundstate occurring
on the first site corresponding to a linear combination
of the vacuum and filled state:
\begin{equation}|0>+ e^{i\chi}|\uparrow ,\downarrow >.\end{equation}
Since the hopping term always turns this Andreev boundstate into
a state with a single electron at site 1 it produces a high energy
state and its effects are therefore suppressed when $|\Delta_B|>>t$.
In the original SN interface model we may think of the Andreev
boundstate as blocking electron transport across the interface and
hence suppressing Andreev scattering.  

This research
 was supported by NSERC of Canada and the NSF under grant No. PHY94-07194.
\begin{appendix}
\section{Integrating out the superconducting electrons}
In this appendix, we outline the steps leading to the correspondence between
the parameters in the interface and boundary models.

Let us thus consider the interface model, which has nonzero gap on sites
$j \leq 0$.  Our strategy will consist in integrating these out.  
For simplicity, let us use the notation $\psi_{1 \sigma}^B$
for the fields on site one.  Omitting all sites with $j > 1$ for ease of
notation, we can write down the contribution to the imaginary-time action
coming from the gapped side and its coupling to the boundary fields:
\begin{eqnarray}
S^+ = \frac{1}{\beta} \sum_{\omega_n} \left\{ \left(\sum_{j \leq 0}
\psi_{j \sigma}^{\dagger} (\omega) [i \omega -\mu] \psi_{j \sigma} (\omega)+
(-t' \psi_{j-1 \sigma}^{\dagger} (\omega) \psi_{j \sigma} (\omega) 
+ \Delta \psi_{j \uparrow}^{\dagger} (\omega) \psi_{j \downarrow}^{\dagger}
(-\omega) +h.c.) \right) - \right. \nonumber \\
\left. (-t'' {\psi_{1 \sigma}^B}^{\dagger} (\omega)\psi_{0 \sigma}(\omega) + 
h.c.) \right\}.
\end{eqnarray}
Fourier transforming as $\psi_{j \sigma} = \frac{2}{\pi}
\int_0^{\pi} dk \sin k(j-1) \psi_{\sigma} (k)$ (for $j\leq 0$) 
and using the Bogoliubov 
transformation
\begin{eqnarray}
\left ( \begin{array}{c} \psi_{\uparrow} (\omega, k) \\
\psi_{\downarrow}^{\dagger} (-\omega, k) \end{array} \right) =
\left( \begin{array}{cc} u(k) & -v^*(k) \\ v(k) & u^*(k) 
\end{array} \right)
\left( \begin{array}{c} \eta_+ (\omega, k) \\ \eta_-^{\dagger} 
(-\omega, k) \end{array} \right)
\end{eqnarray}
where
\begin{eqnarray}
u(k) = \frac{e^{i \chi}}{\sqrt{2}} \sqrt{1 + \frac{\epsilon (k)}
{E (k)}}, \hspace{0.5cm} v(k) = \frac{1}{\sqrt{2}} 
\sqrt{1 - \frac{\epsilon (k)}
{E (k)}}
\end{eqnarray}
and $\epsilon (k) = -2t' \cos k -\mu$ and $E (k) = \sqrt{
\epsilon^2 (k) + |\Delta|^2}$, we arrive at the form
\begin{eqnarray}
S^+ = \frac{1}{\beta} \sum_{\omega_n} \frac{2}{\pi} \int_0^{\pi}
dk \left\{ \eta_{\sigma}^{\dagger} (\omega, k) [i\omega + E(k)]
\eta_{\sigma} (\omega, k) 
[-t''\eta_+^{\dagger} (\omega,k) [u^*(k) \psi_{1 \uparrow}^B (\omega)
- v^*(k) {\psi_{1 \downarrow}^B}^{\dagger} (-\omega) ] \sin k -
\right. \nonumber \\ \left.
-t''\eta_-^{\dagger} (\omega,k) [u^*(k) \psi_{1 \downarrow}^B (\omega)
+ v^*(k) {\psi_{1 \uparrow}^B}^{\dagger} (-\omega) ] \sin k + h.c.] \right\}.
\end{eqnarray}
Integrating out the $\eta$ fields finally gives the boundary action
\begin{eqnarray}
S_B = \frac{1}{\beta} \sum_{\omega_n} \left\{ [i \omega c_1(\omega)
-c_2(\omega)] {\psi_{1 \sigma}^B}^{\dagger} (\omega)\psi_{1 \sigma}^B
(\omega) 
+ (c_1 (\omega) \Delta {\psi_{1 \uparrow}^B}^{\dagger} (\omega)
{\psi_{1 \downarrow}^B}^{\dagger} (-\omega) + h.c.) \right\}
\end{eqnarray}
where the coefficients $c_i (\omega)$, appearing respectively in front
of the pairing-like and potential-like amplitudes, are given by
\begin{eqnarray}
c_1(\omega) = -\frac{4 {t''}^2}{\sqrt{\omega^2 + |\Delta|^2}} 
\Im M^{-2}, \hspace{2.0cm}
c_2(\omega) = -\frac{16 {t'}^2 {t''}^2}{\sqrt{\omega^2 + |\Delta|^2}}
\Im M^{-4} -\mu c_1 (\omega) 
\end{eqnarray}
in which
\begin{eqnarray}
M \equiv \sqrt{-2t' + \mu + i\sqrt{\omega^2 +|\Delta|^2}} 
+ \sqrt{2t' + \mu + i\sqrt{\omega^2 +|\Delta|^2}}.
\end{eqnarray}

The desired correspondence between the interface and boundary parameters
thus takes the form (when $\omega \rightarrow 0$)
\begin{eqnarray}
\Delta_B = \Delta c_1 (\omega \rightarrow 0), \hspace{0.5cm}
V_B = V - c_2 (\omega \rightarrow 0).
\end{eqnarray}
Taking the limit explicitly reproduces equations (\ref{correspondence}).
The frequency-dependent terms are suppressed by powers of $\omega/|\Delta|$
and have thus been ignored at energies well below the gap.  Furthermore,
in the presence of interactions, we expect the above procedure to work
as well, namely that the final result is simply some effective boundary
pairing and scattering potentials.

\end{appendix}


\begin{references}
\bibitem{AndreevZETF} A. F. Andreev, Zh. Eksp. Teor. Fiz. 46, 1823 (1964)
[Sov. Phys. JETP 19, 1228 (1964)]; 49, 655 (1966) [22, 455 (1966)].
\bibitem{DeGennesBook} P.-G. de Gennes, {\it Superconductivity of Metals
and Alloys}, Benjamin, New York, 1966.
\bibitem{DemersPRB4} A. Griffin and J. Demers, Phys. Rev. {\bf B} 4, 
2202 (1971).
\bibitem{BlonderPRB25} G. E. Blonder, M. Tinkham and T. M. Klapwijk, 
Phys. Rev. {\bf B} 25, 4515 (1982).
\bibitem{MaslovPRB53} D. L. Maslov, M. Stone, P. M. Goldbart and D. Loss,
Phys. Rev. {\bf B} 53, 1548 (1996).
\bibitem{Fazio1} R. Fazio, F.W.J. Hekking and A.A. Odintsov, Phys. Rev. {\bf B53},
6653 (1996).
\bibitem{Takane}Y. Takane, J.Phys. Soc. Japan, {\bf 66},
537 (1997).
\bibitem{Fazio2} R. Fazio, F.W.J. Hekking, A.A. Odintsov and R. Raimondi,
cond-mat/9811217.
\bibitem{Kane} C.L. Kane and M.P.A. Fisher, Phys. Rev. {\bf B46}, 15233 (1992).
\bibitem{Deutscher} G. Deutscher and P. Nozi\`eres, Phys. Rev. 
{\bf B50}, 13557 (1994).
\bibitem{Affleck1}For a review see I. Affleck, Acta Physica Polonica, 
{\bf 26}, 1869 (1995); cond-mat/9512099.
\bibitem{Wong} E. Wong and I. Affleck, Nucl. Phys. {\bf B417}, 403 (1994).
\bibitem{ZagoskinJPA30} A. M. Zagoskin  and I. Affleck, J. Phys. {\bf A} 30, 
5743 (1997)
\bibitem{Eggert} S. Eggert and I. Affleck, Phys. Rev. {\bf B46}, 10866 (1992).
\bibitem{Affleck2}I. Affleck, J. Phys. {\bf A31}, 2761 (1998).
\bibitem{Guinea} F. Guinea, Phys. Rev. {\bf B32}, 7518 (1985).
\bibitem{Fledderjohan} A. Fledderjohann, C. Gerhardt, K.-H.. M\"utter, A. Schmitt
and M. Karbach, Phys. Rev. {\bf B54}, 7168 (1996).
\bibitem{Cabra} D.C. Cabra, A. Honecker and P. Pujol, Phys. Rev. {\bf B58}, 6241 (1998).
\bibitem{ChangPRB49} L.-F. Chang and P. F. Bagwell, Phys. Rev. {\bf
B49}, 15 853 (1994).
\end{references}
\end{document}